\documentclass[aip,amsmath,amssymb,reprint,floatfix]{revtex4-1}

\usepackage{graphicx}
\usepackage{dcolumn}
\usepackage{bm}

\usepackage[utf8]{inputenc}
\usepackage[T1]{fontenc}
\usepackage{mathptmx}
\usepackage{etoolbox}
\usepackage{siunitx}

\draft 

\makeatletter
\def\@email#1#2{%
 \endgroup
 \patchcmd{\titleblock@produce}
  {\frontmatter@RRAPformat}
  {\frontmatter@RRAPformat{\produce@RRAP{*#1\href{mailto:#2}{#2}}}\frontmatter@RRAPformat}
  {}{}
}%
\makeatother

\begin{document}


\title{Development of a Scanning Tunneling Microscope for Variable Temperature Electron Spin Resonance} 



\author{Jiyoon Hwang}
  \affiliation{Center for Quantum Nanoscience, Institute for Basic Science, Seoul 03760, South Korea}
  \affiliation{Department of Physics, Ewha Womans University, Seoul 03760, South Korea}
\author{Denis Krylov}
 \affiliation{Center for Quantum Nanoscience, Institute for Basic Science, Seoul 03760, South Korea}
\affiliation{Ewha Womans University, Seoul 03760, South Korea}
\author{Robertus J. G. Elbertse}
\affiliation{Department of Quantum Nanoscience, Kavli Institute of Nanoscience, Delft University of Technology, Lorentzweg 1, Delft 2628 CJ, The Netherlands.}
\author{Sangwon Yoon}
\affiliation{Center for Quantum Nanoscience, Institute for Basic Science, Seoul 03760, South Korea}
\affiliation{Ewha Womans University, Seoul 03760, South Korea}
\author{Taehong Ahn}
 \affiliation{Center for Quantum Nanoscience, Institute for Basic Science, Seoul 03760, South Korea}
 \affiliation{Department of Physics, Ewha Womans University, Seoul 03760, South Korea}
\author{Jeongmin Oh}
 \affiliation{Center for Quantum Nanoscience, Institute for Basic Science, Seoul 03760, South Korea}
 \affiliation{Department of Physics, Ewha Womans University, Seoul 03760, South Korea}
\author{Lei Fang}
 \affiliation{Center for Quantum Nanoscience, Institute for Basic Science, Seoul 03760, South Korea}
 \affiliation{Ewha Womans University, Seoul 03760, South Korea}
\author{Won-jun Jang}
\affiliation{Samsung Advanced Institute of Technology, Suwon 13595, South Korea}
\author{Franklin H. Cho}
 \affiliation{Center for Quantum Nanoscience, Institute for Basic Science, Seoul 03760, South Korea}
 \affiliation{Ewha Womans University, Seoul 03760, South Korea}
\author{Andreas J. Heinrich}\thanks{Corresponding authors}
 \affiliation{Center for Quantum Nanoscience, Institute for Basic Science, Seoul 03760, South Korea}
 \affiliation{Department of Physics, Ewha Womans University, Seoul 03760, South Korea}
\author{Yujeong Bae}\thanks{Corresponding authors}
\email[The authors to whom correspondence may be addressed: ]{bae.yujeong@qns.science and heinrich.andreas@qns.science}
 \affiliation{Center for Quantum Nanoscience, Institute for Basic Science, Seoul 03760, South Korea}
 \affiliation{Department of Physics, Ewha Womans University, Seoul 03760, South Korea}

\date{\today}

\begin{abstract}
Recent advances in increasing the spectroscopic energy resolution in scanning tunneling microscopy (STM) have been achieved by integrating electron spin resonance (ESR) with STM. Here, we demonstrate the design and performance of a home-built STM capable of ESR at temperatures ranging from 1 K to 10 K. The STM is incorporated with a home-built Joule-Thomson refrigerator and a 2-axis vector magnet. Our STM design allows for the deposition of atoms and molecules directly into the cold STM, eliminating the need to extract the sample for deposition. In addition, we adopt two methods to apply radio-frequency (RF) voltages to the tunnel junction, the early design of wiring to the STM tip directly, and a more recent idea to use an RF antenna. Direct comparisons of ESR results measured using the two methods and simulations of electric field distribution around the tunnel junction show that, despite their different designs and capacitive couplings to the tunnel junction, there is no discernible difference in the driving and detection of ESR. Furthermore, at a magnetic field of $\sim$1.6 T, we observe ESR signals (near 40 GHz) sustained up to 10 K, which is the highest temperature for ESR-STM measurement reported to date, to the best of our knowledge. Although the ESR intensity exponentially decreases with increasing temperature, our ESR-STM system with low noise at the tunnel junction allows us to measure weak ESR signals with intensities in the sub-fA range. Our new design of ESR-STM, which is operational in a large frequency and temperature range, can broaden the use of ESR spectroscopy in STM and enable the simple modification of existing STM systems, which will hopefully accelerate a generalized use of ESR-STM.
\end{abstract}




\maketitle 

\section{Introduction}

Magnetic single atoms and molecules are the pinnacle of ever-increasing demands for information storage density \cite{Bogani2008,Natterer2017} and quantum information processing \cite{doi:10.1126/science.1249802,doi:10.1126/science.aay6779}. Reaching the ultimate spatial limit requires the technical capabilities to achieve the single spin sensitivity and control. The scanning tunneling microscopy (STM) provides a particularly high spatial resolution that allows it to image single atoms and molecules, characterize their electronic and magnetic states, and manipulate their positions on a surface \cite{PhysRevLett.49.57,Bian2021}. Despite its exclusively high sensitivity with atomic resolution, the spectral resolution of STM is often limited by temperature due to the thermal Fermi-Dirac broadening of the energy of tunneling electrons \cite{PhysRev.165.821, Song2010, Ast2016, PhysRevLett.107.076804}. Even when the spin systems are cooled down below 1 K temperatures at a high cost, achieving spectral resolution of tens of nano-eV in the conventional spectroscopy of STM, such as inelastic electron tunneling spectroscopy (IETS), is impossible. In contrast, the detection of spin transitions with a few nano-eV energy resolution is easily attainable in electron spin resonance (ESR). Conventional ensemble ESR, however, requires at least $10^7$ spins ($10^{12}$ for nuclear spins) to get strong enough signals\cite{BLANK2003116,Assig2013, Ciobanu2002}, while its spectroscopic resolution is not limited by temperature. The integration of ESR with STM maximizes the powerful benefits of both approaches, and the single spin ESR detection with atomic resolution has been exclusively achieved by ESR spectroscopy in STM \cite{Baumann417,https://doi.org/10.1002/adma.202107534}. 

\begin{figure*}
\includegraphics[width=170mm]{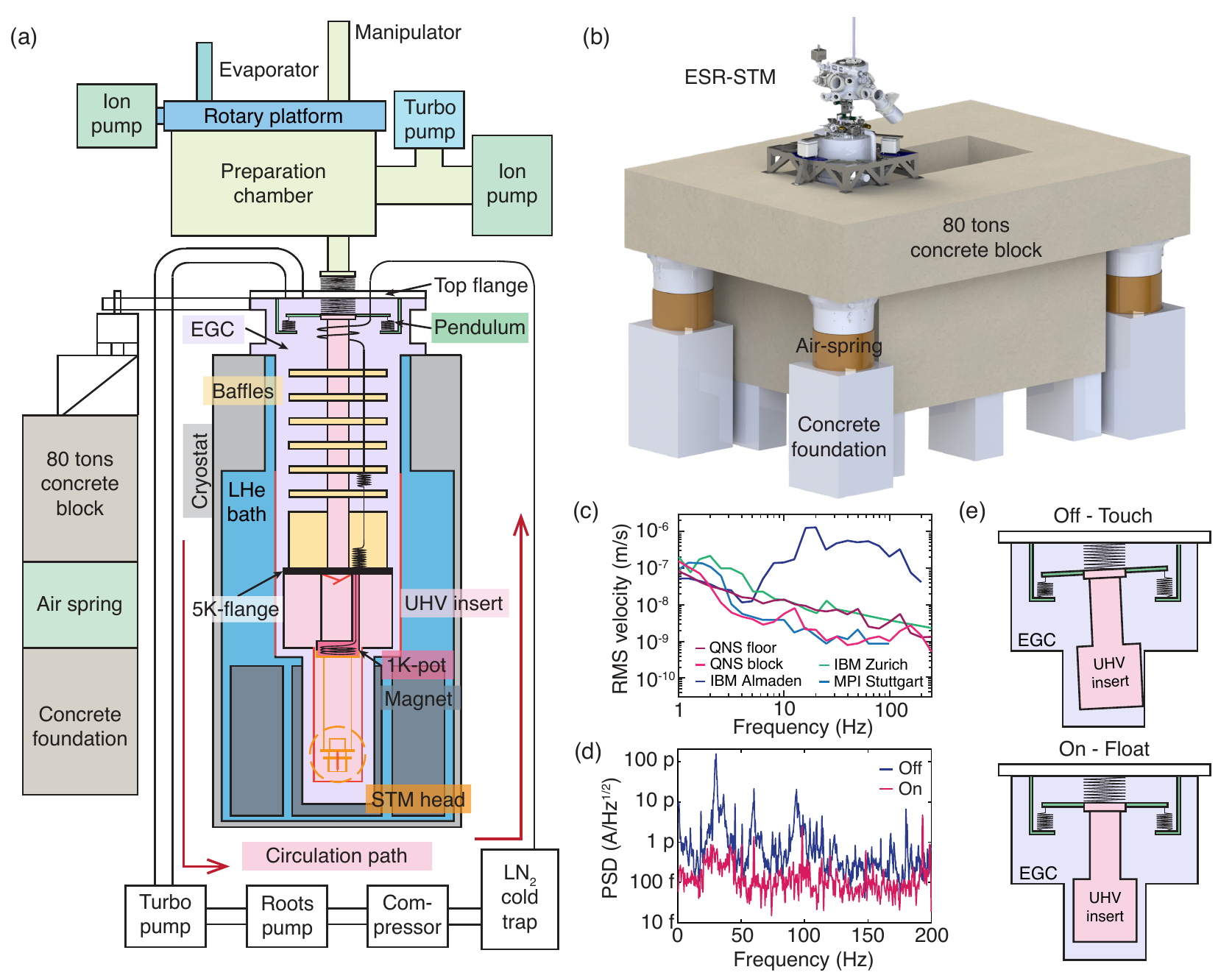}
\caption{\label{fig:structures} Design of the home-built ESR-STM. (a) Overview of the whole system including a cryostat, a vector magnet, a mechanical damping system, a cooling system, an STM, and a preparation chamber. The preparation chamber and the ultrahigh vacuum (UHV) insert are connected to the top flange of an exchange gas can (EGC) via a flexible bellows. The UHV insert is supported by three arms connected to the system via three flexible bellows (labelled as `Pendulum'). (b) ESR-STM setup on a massive concrete block. The cryostat rests on a concrete block floated by four air springs in a sound-proof room. (c) Velocity spectrum measured on the floor of the sound-proof room (purple, QNS floor) and on top of the floating concrete block (magenta, QNS block) in comparison with other precision laboratories (blue, IBM Almaden; green, IBM Zurich; light blue, MPI Stuttgart). (d) Power spectral density (PSD) of the tunnel current measured with pendulum on (magenta) and off (blue) ($V_{\mathrm{DC}}$ = 10 mV, $I$ = 1 nA, Integral-gain = 1 nm/s). (e) Schematics showing pendulum operation. The UHV insert is suspended by three welded bellows, while schematics shows only two bellows for simplicity. When the pendulum is in the `off' state (top), the UHV insert is misaligned and thus in mechanical contact with EGC. In the `on' state (bottom), the pendulum is aligned inside the EGC without touching the EGC and, thus, mechanically isolated from the rest of the system.}
\end{figure*}

ESR measurements in STM require the use of specialized cables for the injection of radio-frequency (RF) electric fields into the tunnel junction \cite{doi:10.1063/1.4955446,PhysRevResearch.2.013032}. However, adding RF cables or replacing existing cables with RF cables to a low-temperature STM for ESR capability is not so trivial since most STMs are already compactly wired and semi-rigid RF cables may cause vibration and heat transmissions to the tunnel junction \cite{doi:10.1063/1.5065384,doi:10.1063/1.5104317,doi:10.1063/1.5109721, Weerdenburg2021, drost2021combining}. Furthermore, to obtain strong ESR signals from a single atom/molecule, it is important to sufficiently polarize the spin state, which imposes a strict requirement on temperature of the system. Given that many existing STMs operate in the 2--4 K temperature range or above, an external magnetic field greater than 1 T facilitates the polarization of surface spins, necessitating RF fields with frequencies higher than 30 GHz at the tunnel junction. 

In this work, we introduce an extremely stable, home-built ESR-STM which operates at temperatures ranging from 1 K to at least 10 K. We use RF cables for the STM tip and an antenna parallel to the sample plane, which enables us to examine RF signal transmission to the tunnel junction through two different cabling schemes. The RF cable going directly to the STM tip allows for a high transmission to the STM junction, where the STM tip apex is positioned within a nanometer scale from an atom or a molecule on a surface. However, the resonance frequency available at the tunnel junction is limited to around 35 GHz due to the requirement for a flexible coaxial cable for the coarse motion of the STM tip. In contrast, for the antenna placed about 4 mm distant from the tunnel junction, we observe that a larger frequency window than the direct wiring to the STM tip is available at the tunnel junction, enabling us to perform high temperature ESR measurements up to 10 K. While the ESR intensities exponentially decrease at elevated temperatures, our ESR-STM enables us to detect weak ESR signals such as sub-fA. Extending the temperature range accessible for the ESR spectroscopy in STM and providing the sensitivity without incurring large costs to modify existing STMs for a lower temperature operation will hasten the spread of this technique for the phase-sensitive control and detection of spins on a surface. 

\section{Instrument design}

We present the design of an ultrahigh vacuum (UHV) ESR-STM operational at a broad range of temperatures with high mechanical stability. The ESR-STM system comprises a cryostat (MAXes\textsuperscript{TM} 6T/4T vector magnet, American Magnetics Inc.), a cooling system, a UHV insert with an STM, and a UHV preparation chamber [Fig.~\ref{fig:structures}(a)]. The liquid \textsuperscript{4}He (LHe) bath cryostat contains a commercially available two-axis vector magnet (American Magnetics Inc., maximum 6 T in-plane and 4 T out-of-plane of a sample) and an exchange gas can (EGC) which is filled with \textsuperscript{4}He exchange gas. The UHV insert, which includes the transfer neck with baffles, `5K-flange', `1K-pot', and STM, is positioned inside of the EGC and cooled by \textsuperscript{4}He exchange gas without direct contact with the LHe reservoir. The 1K-pot contains a home-built Joule-Thomson (JT) refrigerator which serves as the system's coldest location. The STM is rigidly attached to the 1K-pot by materials having high mechanical stiffness and high thermal conductivity. Samples are prepared \textit{in-situ} in a separate preparation chamber and transferred to the STM sample stage using a home-built vertical manipulator with a travel length of about 2 m. In the following sections, we describe each part in detail.

\subsection{Mechanical dampers}

The STM measurement relies on the precise placement of a metallic tip a nanometer apart from the surface \cite{PhysRevLett.49.57,Bian2021}. The extremely high spatial resolution is predicated on the exponential sensitivity of tunnel current on the separation between the STM tip and a sample. For the same reason, ambient vibrations reaching the tunnel junction are exponentially amplified in the tunnel current signals. Constructing high-performance STM with atomic-scale spatial precision, thus, necessitates mitigating the impact of external vibration sources on the tunnel junction. Ambient vibration sources include mechanical vibrations and acoustic noise. 

To avoid such noise sources from being transmitted to the tunnel junction, we installed the entire STM system on a floating concrete block, which weighs 80 tons, in a sound-proof room. The room is located in a low vibration facility built on a 1.4 m thick concrete foundation which is separated from the rest of the building [Fig.~\ref{fig:structures}(a) and (b)]. The massive concrete block is placed in the center of the sound-proof room, separated by about 1 m from the acoustically absorptive wall, and lifted up by four air springs (BiAir-ED-HE-MAX, Bilz Vibration Technology AG) to further isolate the block from floor vibrations. 

Figure~\ref{fig:structures}(c) compares the facility's velocity spectrum to that of other facilities \cite{Loertscher2013,Bian2021}, which was measured using an accelerometer (731A/P31, Wilcoxon) on the floor of the sound-proof room and on the concrete block under ambient vibration conditions. The measured linear acceleration is integrated to obtain linear velocity, which is then converted to an A-weighted one-third octave band spectrum after Fourier transform. The floor noises higher than 3 Hz are significantly suppressed by the massive inertia block, resulting in a one order of magnitude reduction in noise on the block. Given that the lowest fundamental acoustic mode is at 11.16 Hz in the 5.69 m $\times$ 7.68 m $\times$ 6.28 m size room, the feature appearing at around 12 Hz in the noise spectrum [Fig.~\ref{fig:structures}(c), QNS block] is presumably the result of acoustic stimulation of the concrete block motion\cite{MacLeod2016}. The noise level we attained in the overall frequency range relevant to STM performance is comparable to or even better than noise levels measured in the world's leading precision laboratories [Fig.~\ref{fig:structures}(c)]. At considerably higher frequencies, where the STM stage's resonances are located, little environmental vibrations are transmitted. 

While the entire system is rigidly mounted on the floating concrete block, we provide one more damping stage for the STM. To reduce vibration transfer and provide flexibility to adjust alignment, the UHV insert, which includes the STM unit near its lower end, is connected to the top flange using one central bellows for the connection of the preparation chamber and three side bellows for three supporting arms [Fig.~\ref{fig:structures}(a) and (e)]. The UHV insert can be floated by controlling the pressure in the three side bellows and lifting three supporting arms, which suspends the STM like a pendulum. The 1.5 m tall UHV insert has a 1--2 mm gap with the EGC. Mechanical deformation and misalignment, thus, might cause the mechanical contact of the UHV insert to the EGC. Controlling the pressure of the three side bellows allows us to adjust the alignment of the UHV insert with respect to the EGC. The idea for this design stems from the first low-temperature STM built by Don Eigler at IBM \cite{Eigler1990}. As shown by the power spectral density of the tunnel current [Fig.~\ref{fig:structures}(d)], which is measured at a setpoint of $V_{\mathrm{DC}}$ = 10 mV and $I$ = 1 nA using 10\textsuperscript{9} V/A gain preamplifier (SA-606F2, NF corp), we observed significant noise transfer from the environment to the tunnel junction when the pendulum is not adequately operational [upper in Fig.~\ref{fig:structures}(e)]. When the pendulum is well isolated [lower in Fig.~\ref{fig:structures}(e)], the noise level at the tunnel junction is significantly reduced. The combination of the floating concrete block and the internal pendulum results in a remarkable mechanical stability of the system. 

\subsection{Joule-Thomson refrigerator}

\begin{figure}[t]
\includegraphics[width=85mm]{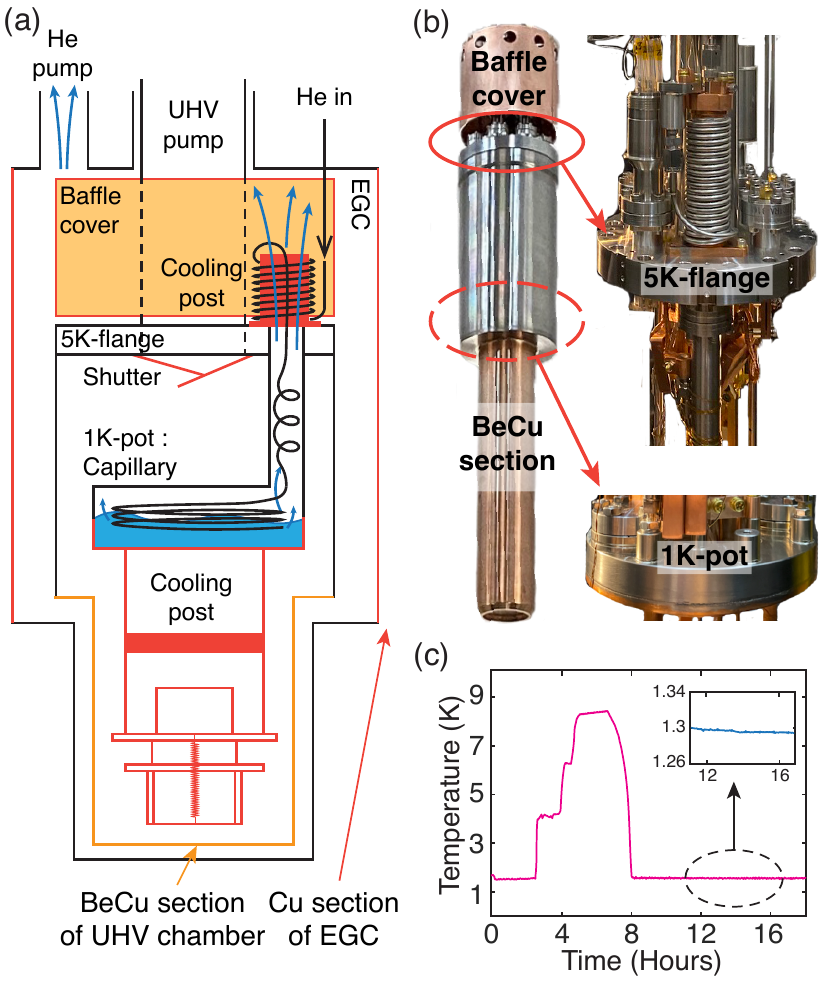}
\caption{\label{fig:JT} Joule-Thomson (JT) refrigerator. (a) Overview of the cooling system including JT refrigerator, 5K-flange, and 1K-pot. For the JT refrigerator, the \textsuperscript{4}He gas is cooled through cooling posts and inserted to the capillary at the 1K-pot. The liquefied He is accumulated to the lower Cu part of 1K-pot, where the whole STM structure is attached by Cu rods. The 1K-pot is pumped through the EGC. At the pumping port, the cooling post and baffle cover are located, which cool the He gas and electrical feedthroughs, respectively. (b) Photographs of the UHV insert with baffle cover on 5K-flange, the UHV chamber made of stainless steel and BeCu, 5K-flange with feedthroughs and a cooling post, and 1K-pot. (c) Temperature of STM controlled by a heater installed on the 1K-pot during the operation of JT refrigerator. By adjusting the heating power, the system can stably operate at different temperatures. Inset: STM temperature during the operation of JT refrigerator without heating the STM stage, which can run continuously at 1.3 K.}
\end{figure}

The LHe bath cryostat, containing the superconducting magnets, provides a LHe reservoir (maximum 133 L of LHe) with a hold time of around 140 hours allowing for the operation of magnets. The EGC is housed in the cryostat and is in direct contact with LHe. The EGC is composed of three different segments [Fig.~\ref{fig:structures}(a)]. The top part from room temperature (RT) top flange to the end of the cryostat's neck is made of stainless steel with a small wall thickness ($\sim$7 mm) to minimize thermal conduction. The choice of materials for the middle section is oxygen-free, high conductivity (OFHC) copper (Cu) for better cooling of the UHV insert regardless of the level of LHe in the cryostat. The last part is the tail of the EGC, which is placed in the magnet's center. To avoid the creation of eddy currents while ramping the magnet, the tail is made of stainless steel. The EGC is filled with He exchange gas (0.01--3 mbar). The work shown in this manuscript uses \textsuperscript{4}He as exchange gas, which can be replaced with \textsuperscript{3}He to achieve lower base temperatures.

The UHV insert is situated inside the EGC and cooled by the exchange gas without direct contact with LHe, which lowers the vibration transfer stemming from the boiling of LHe and external vibration sources. The UHV insert includes the transfer neck with baffles, 5K-flange, 1K-pot, and STM [Fig.~\ref{fig:structures}(a) and \ref{fig:JT}(a)]. The transfer neck is a 750 mm long nipple with DN35CF, made of stainless steel, with an outer diameter of 40 mm and a wall thickness of 1.4 mm. We installed six baffles outside the transfer neck with a diameter close to the inner diameter of the EGC, to prevent thermal radiation from reaching the bottom of the UHV insert and to provide a thermal gradient from the RT top flange to the 5K-flange. The baffles are made of gold-plated Cu without Ni plating. All six baffles are located in the top stainless steel part of the EGC except the last one, which is located in the Cu section of the EGC. The 5K-flange maintains a gap of about 2 mm to the EGC, allowing the exchange gas to effectively transfer heat between the EGC and the 5K-flange. All electrical feedthroughs are installed on the 5K-flange to cool electrical cables before entering the STM stage. We additionally installed a baffle cover made of OFHC Cu to help cooling all feedthroughs on the 5K-flange [labelled as `Baffle cover' in Fig.~\ref{fig:JT}(a) and (b)].

To achieve STM operation at temperatures below 5 K, the UHV insert is equipped with a JT refrigerator in the 1K-pot. The 1K-pot is made of two different materials: stainless steel for the upper section and Cu for the lower section [Fig.~\ref{fig:JT}(b) lower right corner]. The 1K-pot with a volume of $\sim$0.1 L is mounted on the UHV side of the 5K-flange above the magnet and thermally isolated from the surrounding UHV chamber. The 1K-pot is continuously fed with cold, compressed He gas through a thin capillary [Fig.~\ref{fig:JT}(a)]. Through the JT effect, some of this gas is liquefied in the 1K-pot. The liquefied He is accumulated in the lower Cu part of the 1K-pot, where the STM is rigidly attached through three rods made of OFHC Cu. The capillary inside the 1K-pot is 0.3 m long and is coiled three times and partially submerged in the LHe in the 1K-pot [Fig.~\ref{fig:JT}(a)]. The capillary is composed of Inconel tube with an inner diameter of 500 $\mu$m. We create a flow impedance by fitting a tungsten wire with a nominal diameter of 500 $\mu$m inside the Inconel tube. In steady-state flow, the phase transition of He happens via the JT effect in the 1K-pot, liquefying some of the exchange gas and making the 1K-pot the coldest place in the entire system \cite{doi:10.1063/1.3520482}. We installed a heater on the 1K-pot to control STM temperatures (see Table \ref{tab:table1} for wiring). As shown in Fig.~\ref{fig:JT}(c), we can accurately and reliably adjust the STM temperature by varying the power applied to the heater. 

The boil-off from the LHe in the 1K-pot goes through a pumping line (a 0.15 m long nipple with DN16CF), where it cools the incoming gas in a counter-flow heat exchanger [Fig.~\ref{fig:JT}(a)]. We used a customized gasket with a reduced inner diameter (12 mm) to avoid additional noise production caused by pumping liquefied He at around the $\lambda$-point of He ($\sim$2.18 K). The pumped gas then exits into the EGC at the 5K-flange. The exchange gas (\textsuperscript{4}He) is pumped out at the RT side using a turbo pump (HiPace 450, Pfeiffer Vacuum) and a roots pump (A100L, Pfeiffer Vacuum), compressed using a customized diaphragm pump (N186.1.2AN/V.12E, KNF), cleaned by a cold trap, and injected back into the capillary located at the 1K-pot [Fig.~\ref{fig:structures}(a)]. Prior to entering the 1K-pot, the compressed gas passes two precooling stages, on one of baffles and the 5K-flange, where the 1/8 inch diameter stainless steel tube for the insert gas is brazed around Cu rods [Fig.~\ref{fig:JT}(a) and (b)]. Continuous operation of the JT refrigerator has no perceptive effect on the hold time of LHe in the cryostat or the noise level at the tunnel junction, which allows us to keep the JT refrigerator running during STM operation. To minimize heat input from thermal radiation, the 1K-pot and STM are protected by a radiation shutter at the end of the transfer neck [Fig.~\ref{fig:JT}(a)]. 

\subsection{Design of RF antenna}

\begin{figure}
\includegraphics[width=85mm]{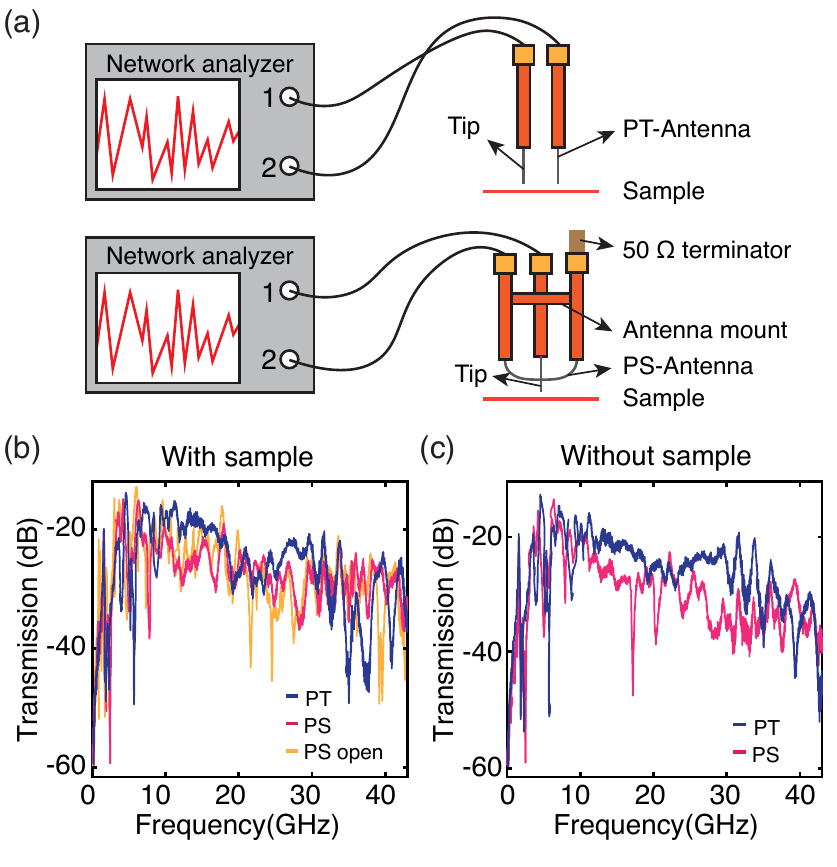}
\caption{\label{fig:antenna} RF signal transmission of two different antennas. (a) Experimental setup for RF transmission test of the PT- (upper) and PS-antennas (lower). Using a network analyzer, RF signals are applied to the antenna and the transmitted signals to the tip are measured. The PT- and PS-antennas are made of stripped parts of coaxial cables which run parallel to the tip and to the sample, respectively. The PS-antenna is terminated with a 50-Ohm load and held in place with the help of an antenna mount. (b) Signal transmission for the PT-antenna (blue) and PS-antenna with (magenta) and without (yellow) a 50-Ohm terminator. (c) Signal transmission for the PT-antenna (blue) and PS-antenna (magenta) without the sample. While the transmission for the PT-antenna remains nearly identical, we observed a considerable loss in transmission between the PS-antenna and the tip when the sample was removed. 
}
\end{figure}

The first demonstration of ESR of single atoms on a surface was achieved by applying RF voltages into the tunnel junction via the STM tip (referred to as the "original ESR-STM design"), where the RF voltages are combined with the DC bias voltage at RT using a bias tee\cite{Baumann417, doi:10.1063/1.4955446}. Due to the complex modifications to the RF cablings of STM in the early design, Seifert et al.\cite{PhysRevResearch.2.013032} introduced one simple antenna capacitively coupled to the tip, where the antenna was installed as parallel as possible to the tip (PT-antenna). The PT-antenna offers larger signal transmission to the tunnel junction at high frequency than the original ESR-STM design. Here, we introduce a new design of an antenna parallel to the sample (PS-antenna) and compare two different antennas (PT- and PS-antennas) by measuring and simulating their signal transmission to the tip and simultaneously to the sample. We provide a universal approach to integrating RF cables into the STM, while the antenna design was dedicated based on the geometry of our STM by considering three aspects: i) stable transmission while allowing for the coarse motion of the tip, ii) mechanical stability, and iii) suppression of standing waves.

Figure~\ref{fig:antenna}(a) shows the experimental setup for the characterization of signal transmission from two different antennas (PT- and PS-antennas) to the tip-sample junction using a network analyzer (N5224B, Keysight). We set the tip and each antenna using semi-rigid coaxial cables (UT-085C-LL, Carlisle Interconnect Tech.) and the sample using a Cu foil. The PT-antenna and the tip are made of coaxial cables with unshielded inner conductors at the ends, while the PS-antenna is made by stripping the outer conductor in the middle of a coaxial cable and terminating it with a 50-Ohm load [Fig.~\ref{fig:antenna}(a)]. We used an antenna mount to hold two sides of the PS-antenna’s outer conductors to rigidly support the antenna and thus provide better mechanical stability [lower in Fig.~\ref{fig:antenna}(a)]. Note that the metallic antenna mount may behave like a signal receiver when it is less than 50 mm away from the antenna. The tip and each antenna are connected to the network analyzer for transmission (S21) measurements using flexible cables (FLC-1M-SMSM+, Mini-Circuits), and calibrations are performed using an electronic calibration module (Ecal N4693A (4693-60003), Keysight) to eliminate the systematic errors from the flexible cables.

Despite their different designs, both PT- and PS-antennas exhibit a uniform decrease over the whole frequency range, with sharp dips at frequencies lower than 10 GHz. We ascribe such sharp attenuations to interferences of incident and reflected signals at unshielded parts of the semi-rigid cables. The transmission of an antenna often has a strong frequency dependence where the resonance frequency is determined by its size and shape. RF signals with wavelengths longer than the dimensions of an antenna are difficult to transmit through capacitive coupling\cite{RFandMICROWAVE_ENGINEERING}. Such losses at low frequencies remained even after we removed the sample [Fig~.\ref{fig:antenna}(c)], supporting our hypothesis. 

While the signal transmissions through two different antennas are comparable, we found distinct transmission paths between two antennas. The shape and dimension of the PT-antenna are chosen to be similar to those of the tip and, thus, we expect maximum transmission between the two by a capacitive coupling as the antenna is placed parallel to the tip. In the range of 10 to 30 GHz, we indeed observe less loss from the PT-antenna compared to the PS-antenna [Fig.~\ref{fig:antenna}(b)]. The PS-antenna, on the other hand, is less coupled to the tip since it is designed to be parallel to the sample and thus practically perpendicular to the tip. We also note that there is a considerable change in the transmission for the PS-antenna when the sample was removed, indicating significant coupling of the PS-antenna to the sample [Fig.~\ref{fig:antenna}(b) and (c)]. Despite rather drastic changes in the design, these results qualify that the PS-antenna is an adequate alternate to the PT-antenna, particularly at frequencies above 30 GHz as we see better transmission properties. Note that we have a 20 mm long unshielded part of the cable for the PT-antenna. Adjusting the stripped area possibly provides additional room to improve the signal transmission.

We briefly discuss the role of the 50-Ohm terminator in the PS-antenna. A proper termination of an RF cable is essential to reduce the creation of standing waves. Ideally, if the load impedance at the cable’s end is matched to the characteristic impedance of the transmission cable, no reflections appear\cite{RFandMICROWAVE_ENGINEERING}. However, in reality, the impedance mismatch between the cable (50 Ohm) and the STM junction (typically 1 GOhm) causes significant reflections and thus the creation of standing waves. To suppress reflections, we put a 50-Ohm terminator (BM11132, Bracke) rated up to 40 GHz at the end of the transmission cable [Fig.~\ref{fig:antenna}(a)]. Figure~\ref{fig:antenna}(b) shows signal transmissions from the PS-antenna to the tip with and without the 50-Ohm terminator, designated as PS and PS open, respectively. We observed that adding the 50-Ohm terminator helps reducing some abrupt features in the transmission.

\begin{figure}
\includegraphics[width=85mm]{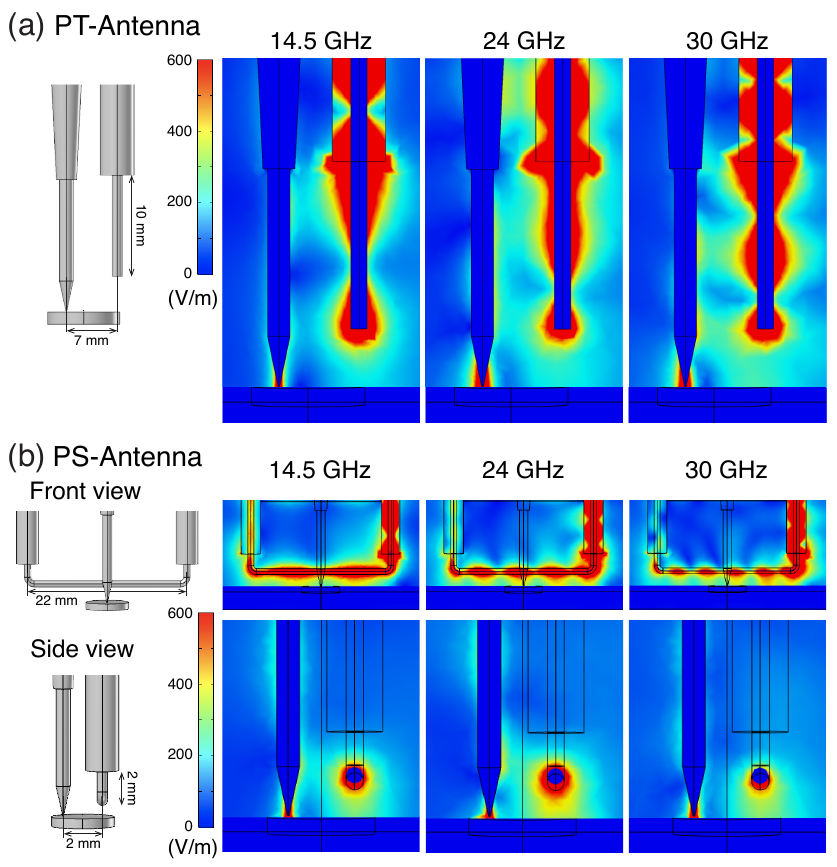}
\caption{\label{fig:COMSOL} COMSOL simulations of electric field distribution. (a) Schematics of the simulation setup and the simulated images of electric fields around the antenna, tip, and sample for the PT- and (b) PS-antenna when oscillating voltage signals are applied through the antennas. Three selected simulation results are shown when the frequency of the incident oscillating voltage is set to 14.5, 24, and 30 GHz.}
\end{figure}


The signal transmissions of each antenna to the tip-sample junction are further investigated using COMSOL multiphysics to simulate the electric field in space. In the simulation, we provide an oscillating voltage of 1 V to the transmission line of the antenna, which is then transmitted to the junction. Figures~\ref{fig:COMSOL}(a) and (b) show simulated results calculated at incident electric field frequencies of 14.5, 24, and 30 GHz for the PT- and PS-antennas. The simulation results reveal that for both PT- and PS-antennas a strong localized electric field is generated at the tip-sample junction. 
On average, the PS-antenna transmits somewhat less to the junction than the PT-antenna across all frequencies (5--35 GHz). However, we observed that at some frequencies (for example at 14.5 GHz), the PT-antenna barely transmits signals. Because the tip and the PT-antenna have similar dimensions, the transmission might change drastically at particular frequencies where a half of a wavelength ($\sim$10 mm) is comparable to the length of the PT-antenna or the tip. The PS-antenna, on the other hand, is less frequency-sensitive due to its highly distinct dimensions from the sample. Furthermore, the PS-antenna partially transmits signals through the tip, which allows it to transmit the RF signals without considerable frequency dependence by transmitting the signals through alternative paths. We note that moving the tip laterally along the antenna results in no significant change of the localized electric field at the tip-sample junction.
The simulation results support the practical replacement of the PT-antenna with the PS-antenna, particularly for STMs where the sample is fixed while the tip moves in large coarse steps. In the later sections, we will demonstrate that this newly-designed antenna is indeed competitive with the PT-antenna for ESR-STM measurement.
\begin{figure}
\includegraphics[width=85mm]{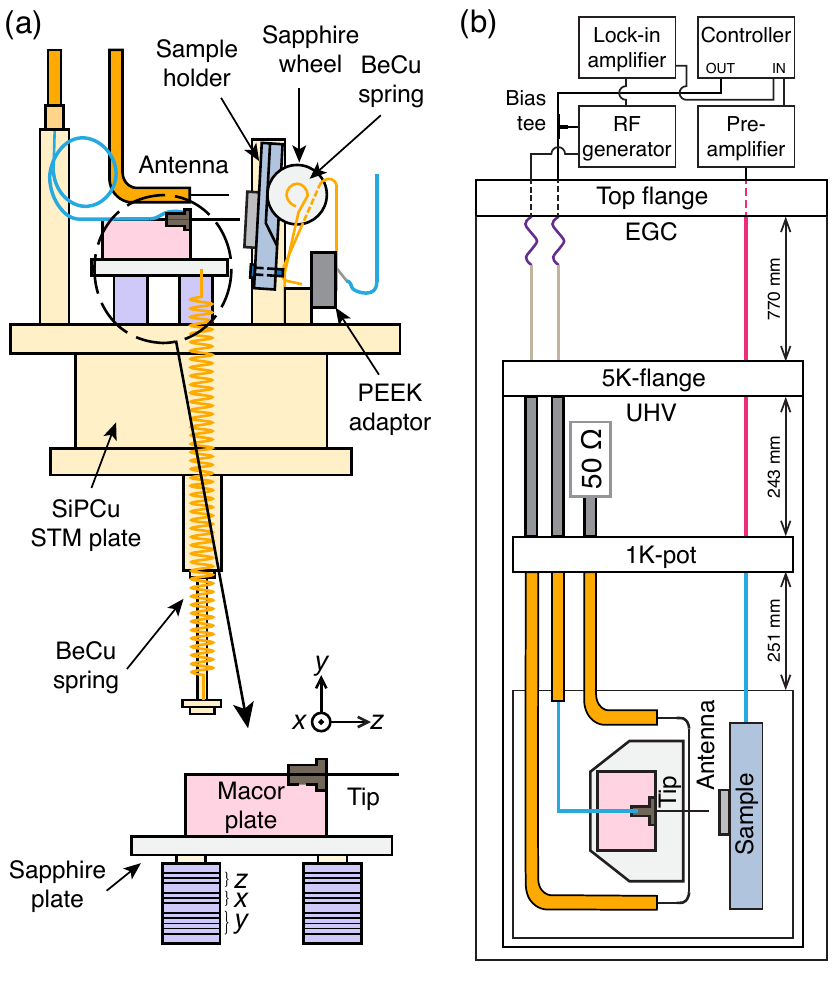}
\caption{\label{fig:wiremap} STM stage and cabling schemes. (a) Overview of the STM stage including the tip, antenna, sample stage, and piezo stacks. Coaxial cables (orange and cyan) are used for connections to the tip and the antenna. The sample holder is clamped by BeCu springs (light orange) with sapphire wheels (light gray). A molybdenum (Mo) bolt of the sample holder is in contact with the BeCu spring (light orange), which is connected to the inner conductor of a flexible coaxial cable (cyan) as a tunnel current signal line. The PtIr tip is in a tin plated BeCu pin spring socket (2-330808-7, AMP-TE connectivity) as a tip holder which is mounted in the Macor tip plate on the sapphire plate sitting on three piezo stacks for the tip's coarse and scan motions (below). Only two stacks are shown in the side view. (b) Wiring schemes for ESR-STM measurement. The same RF cable schemes are used for the tip and antenna except for the cables between the 1K-pot to STM stage. For the tip side, a flexible coaxial cable is used between the semi-rigid Cu cable and the tip. For the antenna, the semi-rigid Cu cable is partially stripped near the sample, then returned back to the 1K-pot, and followed by NbTi cable with a 50-Ohm terminator.}
\end{figure}

\subsection{Scanning tunneling microscope with RF cabling}
\begin{table*}
\caption{\label{tab:table1}Summary of electrical cables used in the system}
\begin{tabular}{|m{2.4cm}||m{4.7cm}|m{4.7cm}|m{5.2cm}|}
\hline
\textbf{ } & \textbf{RT top flange to 5K-flange} & \textbf{5K-flange to 1K-pot} & \textbf{1K-pot to STM}\\
\hline 
\textbf{Piezo stacks} & CC-SS, Lake Shore Cryotronics & CC-SS, Lake Shore Cryotronics & TYP3-2TW-15, Accu-Glass Products\\
\hline
\textbf{Sample current} & N12-36M-131, New England Wire & N23-44M-100-0, New England Wire & CW-2040-3650P, Cooner Wire\\
\hline
\textbf{Tip bias} & SCA49141-09, Fairview; & SC-160/50-NbTi-NbTi, Coax Japan & UT-085C-LL, Carlisle Interconnect Tech;\\
& SC-119/50-SSS-SS, Coax Japan &  & CW2040-3650P, Cooner Wire\\
\hline
\textbf{Antenna} & SCA49141-09, Fairview;& SC-160/50-NbTi-NbTi, Coax Japan & UT-085C-LL, Carlisle Interconnect Tech\\
& SC-119/50-SSS-SS, Coax Japan& & \\
\hline
\textbf{Temperature} & Silicon diodes / Cernox& Cernox & \\
\textbf{Sensors}& with Quad-Lead$\textsuperscript{TM}$ cryogenic wires, &with Quad-twist$\textsuperscript{TM}$ cryogenic wire,& -\\
& Lake Shore Cryotronics&Lake Shore Cryotronics& \\
\hline
\textbf{Heaters}&WMW-32, Lake Shore Cryotronics&SS 304 insulated with HNY,& -\\
& &California Fine Wire company& \\
\hline
\end{tabular}
\end{table*}

Our STM is relatively simple in design [Fig.~\ref{fig:wiremap}(a)] and comprises the tip, sample stage, and cables for RF signal transmission, electrical control, and STM signal. Our sample is vertically loaded and held in place by two BeCu springs with sapphire wheels [Fig.~\ref{fig:wiremap}(a)]. The sapphire wheels allow us to transfer samples without producing excessive friction and hold the sample in the STM stage without electrical contact to the stage. The sample holder is electrically isolated from the STM stage made of silicon phosphorus copper (SiPCu) by sapphire pieces between them. These sapphire pieces are glued to the rest of STM stage using epoxy (T7110, EPO-TEK). For better thermal coupling, the bottom part of sample holder is contacted to the SiPCu stage without sapphire but covered by a Kepton tape for electrical insulation.

While the sample stays in place, we move the tip for both coarse and fine motion. The coarse motion of the tip is given in two directions, parallel ($x$) and perpendicular ($z$) to the sample surface for lateral coarse positioning and coarse approaching, respectively [Fig.~\ref{fig:wiremap}(a)]. For each direction, we used two shear piezoelectric actuators. The same shear piezo actuators are used for the fine scanning motions in $x$ and $z$. For fine motion in the other in-plane direction ($y$), we used four linear piezo actuators. The total 8 piezo actuators (PAXZ+0063, PI) are thus stacked together, where an alumina plate is inserted between actuators responsible for motion in different directions to prevent electrical interference. A sapphire plate is mounted on top of three sets of these piezo stacks [Fig.~\ref{fig:wiremap}(a)]. A tip mount made of Macor is glued to the sapphire plate and at present, the tip cannot be changed \textit{in-situ}. The tip is made of PtIr wire. If the tip is contaminated and, thus, the tunnel current does not show an exponential dependence on the tip-sample distance, we clean the tip by applying a high bias voltage ($\sim$100 V) between the tip and the sample as part of field emission process while using a preamp with low gain ($10^7-10^8 $ V/A). To protect the STM wire from damage caused by excessively high current flow, we insert a resistor between the high voltage power supply and the tip during field emission. A 50 mm long BeCu spring pulls the sapphire plate against the piezo stacks. The spring is designed in such a way that the normal force exerted by the spring results in right amount of friction for the coarse motion of the slip-stick mechanism without mechanical instability during scanning.

The electrical cables used for STM operation are carefully selected in consideration of the electrical noise transfer, signal loss, and heat transfer for each section [Fig.~\ref{fig:wiremap}(b)]: i) from RT top flange to 5K-flange, ii) from 5K-flange to 1K-pot, and iii) from 1K-pot to STM. The choice of cables for each component is summarized in Table \ref{tab:table1}. From the RT flange to the 5K-flange, all cables are thermalized by the exchange gas in the EGC. For further thermalization, flexible cables are wrapped around gold-plated Cu posts located at baffles. All cables are connected to the UHV side using electric feedthroughs. Since unshielded parts of high voltage cables can cause an electric arc in the EGC filled with He gas at a pressure of 0.01–3 mbar, we paid special attention to the high voltage multi-pin feedthroughs (9132004, MDC) at the RT top flange a
nd the 5K-flange. To protect high voltage cables from the electric arc, we used Teflon tubes to cover the cable ends and connectors, and a hand-strippable coating compound (TURCO5580-G, AeroBase Group) to cover the metallic part of feedthroughs between pins. 

We used coaxial cables both for the tip bias line and for the antenna to inject RF signals to the tunnel junction [Fig.~\ref{fig:wiremap}(b) and Table \ref{tab:table1}]. We used stainless steel semi-rigid coaxial cables (SC-119/50-SSS-SS, Coax Japan) between the RT top flange and the 5K-flange [Fig.~\ref{fig:wiremap}(b) light gray]. The cables are connected to the RT top flange using hand-formable coaxial cables (SCA49141-09, Fairview) [Fig.~\ref{fig:wiremap}(b) purple] to permit pendulum adjustment. Through electrical feedthroughs (242-SMAD27G-C16, Allectra), the cables are connected to semi-rigid NbTi coaxial cables [Fig.~\ref{fig:wiremap}(b) gray] on the UHV side. From 1K-pot, we used different cabling schemes for the tip and antenna. For both, we installed semi-rigid Cu coaxial cables [Fig.~\ref{fig:wiremap}(b) orange]. For the tip, the Cu cable is connected to the PtIr tip by a flexible Cu cable (CW2040-3650p, Cooner Wire), using a tin-plated BeCu pin spring socket (2-330808-7, AMP-TE connectivity) as an adapter. While the RF cable for the STM tip terminates with an open end at the STM stage, the RF cable for the antenna passes the STM stage parallel to the sample surface and terminates with a 50-Ohm terminator near the 5K-flange [Fig.~\ref{fig:wiremap}(b)]. Note that in an earlier work \cite{PhysRevResearch.2.013032}, an antenna was positioned parallel to the tip, which is not ideal in our design due to changing distance between the antenna and the tunnel junction during coarse motion of the tip. Instead, we installed an antenna parallel to the sample plane as introduced in the previous section (Fig.~\ref{fig:antenna}). For the antenna, the outer conductor of the coaxial cable is stripped off by $\sim$20~mm around the tip and the exposed inner conductor acts as the antenna to transfer RF signals to the STM junction. The inner conductor is $\sim$ 2 mm above the tip and $\sim$ 3 mm away from the sample, ensuring that the RF cable does not interrupt the tip's coarse motion or sample transfer. Both sides of the stripped cable are rigidly fixed, which makes the antenna robust against mechanical vibration.

\subsection{UHV chamber and sample preparation}

In our ESR-STM system, the preparation chamber and the UHV insert are in UHV. The preparation chamber is equipped with several e-beam and thermal evaporators, a heating stage, a manipulator, a sputter gun (IQE 11/35, SPECS), an Auger electron spectrometer (ESA 100, STAIB), leak valves for inserting gases (such as Ar, O\textsubscript{2}, and N\textsubscript{2}), a residual gas analyzer (RGA100, Stanford Research Systems), a turbo molecular pump (HiPace 700 H, Pfeiffer Vacuum), and an ion pump (VacIon Plus 500 Starcell, Agilent). In the preparation chamber, we prepare atomically clean substrates and grow ultra-thin insulating layers on metallic substrates, e.g., MgO(100) on Ag(100). After the sample preparation, the sample is transferred from the preparation chamber to the STM stage in the UHV insert using a manipulator with the travel length of $\sim$2 m.

Access to the UHV insert from the preparation chamber is given by shutter mechanisms. There is no visual access to the scan head during the sample transfer procedure. To ensure a safe sample transfer, we installed a transfer guiding ring near the STM stage, which guides the manipulator to reach the STM stage. furthermore, we are able to adjust the pendulum to align the UHV insert (thus, the STM stage) with the manipulator. We have indicators which monitor the movement of the pendulum. During the sample transfer, the indicators show any change of pendulum positions caused by misalignment between the manipulator and the STM stage. 

Our UHV insert does not require bake-out and is pumped through the preparation chamber before cooling the system. Once we start cooling the UHV insert and its pressure becomes lower than the preparation chamber due to cryogenic pumping effect, we close the gate valve between the preparation chamber and the UHV insert. We observed no perceptible degradation of the clean sample for at least 5 months. 

Many STMs require samples to be taken out in order to deposit atoms or molecules on surfaces. This leads to an unavoidable heating of the samples, which often causes unwanted diffusion of adsorbates. It can also lead to intrusion of hydrogen into the STM stage. Our STM design allows direct deposition of atoms and molecules onto the sample surface without removing the sample from the STM stage. As shown in Fig.~\ref{fig:structures}(a), we installed evaporators on the rotary platform. While the manipulator is aligned with the UHV insert during the sample transfer, we rotate the platform to align evaporators with the STM stage for the evaporation of atoms or molecules. Our sample holder made of molybdenum (Mo) is machined to have an $\sim8^{\circ}$ slope. This ensures that some atoms, which are evaporated in the preparation chamber at a distance of $\sim$2 m from the sample, land on the sample surface. Together with a mechanically actuated shutter located at the 5K-flange [Fig.~\ref{fig:JT}(a)], this permits \textit{in-situ} deposition of atoms/molecules on our sample which remains at a temperature of about 10 K during deposition.

\section{System operation and performance characterization}

Using this home-built ESR-STM, we implement ESR spectroscopy of hydrogenated titanium (TiH) atoms on two monolayers (MLs) of MgO on Ag(100) \cite{doi:10.1103/PhysRevLett.122.227203, Baeeaau4159}. 
We characterize frequency-dependent transmission to the tunnel junction through the tip and the newly introduced antenna (Fig.~\ref{fig:wiremap}), which allows us to compare available frequency windows and transmission efficiency. We perform the side-by-side comparison of the two ESR measurements using the same tip and atomic species and ultimately present ESR spectra measured at elevated temperature (10 K) without losing energy resolution. 

\subsection{Transfer function}

\begin{figure}
\includegraphics[width=85mm]{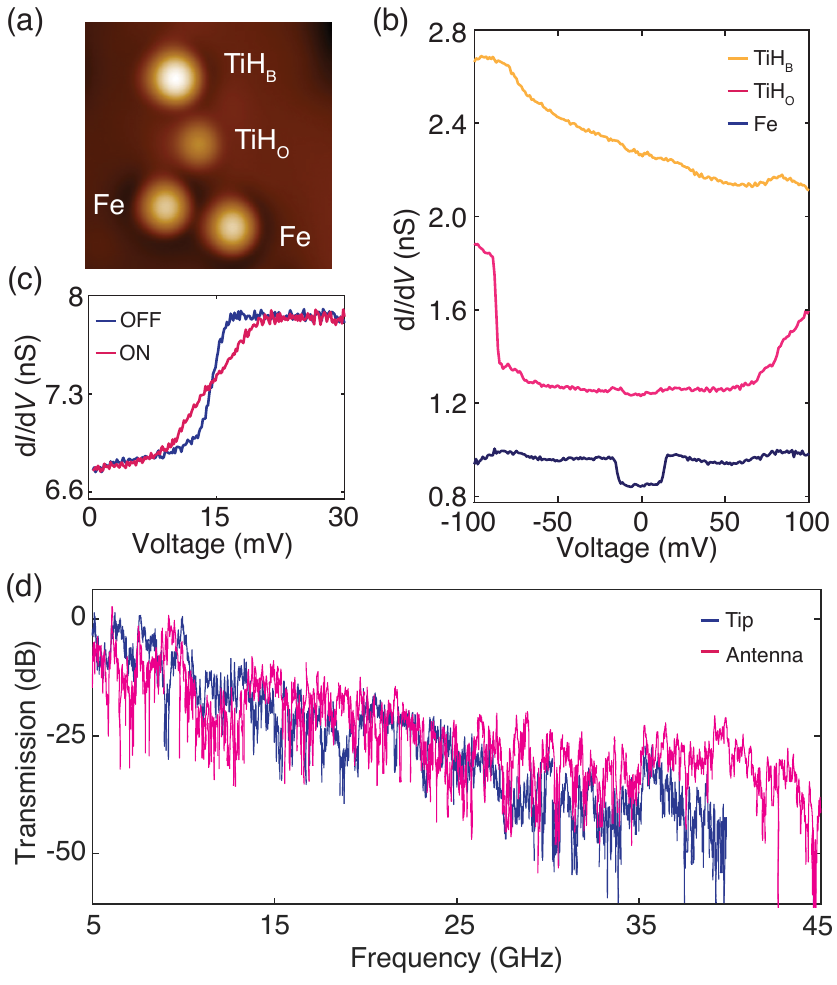}
\caption{\label{fig:tf} Experimental setup for ESR measurement. (a) A constant current STM image of Fe and TiH on MgO/Ag(100) ($V_{\mathrm{DC}}$ = 100 mV, $I$ = 20 pA, T = 1.3 K). (b) Characteristic STM d$I$/d$V$ spectra of Fe and TiH. The Fe and different TiH species (TiH\textsubscript{O}/TiH\textsubscript{B}) are easily identified by their distinct d$I$/d$V$ spectra and apparent heights. Successive curves are shifted by 0.4 and 1.2 nS, respectively, for clarity ($V_{\mathrm{DC}}$ = 100 mV, $I$ = 100 pA, $V_{\mathrm{mod}}$ = 1 mV, $T$ = 1.3 K). (c) Quantifying RF transmission to the tunnel junction. The IETS step of Fe is broadened while applying RF voltages, which is used to quantify the effective RF amplitudes at the tunnel junction ($V_{\mathrm{DC}}$ = 14.3 mV, $I$ = 100 pA, $V_{\mathrm{mod}}$ = 1 mV, $T$ = 1.3 K, $f$ = 16.21 GHz, $P_{\mathrm{RF}}$ = --16 dBm). (d) Transfer functions measured by applying RF voltages through the tip (blue) and the antenna (magenta). Both curves are shifted by --50 dB to have 0 dB correspondent to the zero attenuation of the transmission line.
}
\end{figure}

Figure~\ref{fig:tf}(a) shows a constant current STM image of Fe and TiH on 2 MLs of MgO on Ag(100). We deposited Fe and Ti \textit{in-situ} while the sample was kept below 10 K. The Ti tends to be hydrogenated during the deposition by the residual hydrogen in the UHV chamber. The thickness of the MgO layer was determined from the point-contact measurement and IETS steps of Fe\cite{10.1038/nphys3965}. Spin-polarized tips were made by picking up Fe atoms. TiH, as the simplest spin system\cite{doi:10.1103/PhysRevLett.122.227203, Baeeaau4159}, is chosen for the ESR measurement. While Fe absorbs on top of the oxygen atoms in the MgO lattice, TiH is found at two different binding sites, a bridge between oxygen atoms (TiH\textsubscript{B}) \cite{Baeeaau4159} or atop oxygen atoms (TiH\textsubscript{O}) \cite{doi:10.1103/PhysRevLett.122.227203}. The d$I$/d$V$ spectra [Fig.~\ref{fig:tf}(b)] and apparent heights in constant current STM images [Fig.~\ref{fig:tf}(a)] readily identify Fe and TiH on MgO/Ag(100). 

As a crucial step towards frequency-sweep ESR measurements, we characterize the frequency-dependent transfer function\cite{doi:10.1063/1.4955446} of both RF transmission lines (the tip and the antenna) using IETS steps of Fe at around 14.5 meV [Fig.~\ref{fig:tf}(c)]\cite{10.1038/nphys3965}. Note that IETS steps of TiH\textsubscript{O} are also available\cite{PhysRevResearch.2.013032,drost2021combining,Weerdenburg2021}. The IETS steps are broadened while applying RF power to the tunnel junction. The IETS step broadening, which follows an arcsine distribution, is used as a reference to convert the effective RF voltage at the tunnel junction to a voltage measured by a lock-in amplifier (SR865A, SRS). The lock-in amplifier detects variations in effective RF voltages in the tunnel junction caused by frequency-dependent transmission. Figure~\ref{fig:tf}(d) shows transfer functions in the frequency range of 5--45 GHz. While the transfer functions of the two cabling schemes are comparable over a wide frequency range, the prominent difference is manifested at frequencies higher than 37 GHz and frequencies lower than 15 GHz. Due to the existence of a flexible cable (CW2040-3650p, Cooner Wire) for coarse movement of the tip and a bare wire as the tip, considerable signal loss occurs at frequencies greater than 37 GHz. This limitation is circumvented by using an antenna, which does not require the use of lossy lines and thus enables for transmission at frequencies greater than 37 GHz. At frequencies lower than 15 GHz, however, the antenna transmits less than the tip, which is presumably related to resonances of the antenna ($\sim$20 mm in length) and its surroundings, such as the sample ($\sim$5 mm in diameter) and the tip ($\sim$10 mm in length). 

Based on the Fourier analysis of transfer functions in Fig.~\ref{fig:tf}(d), we found no prominent standing waves for both transmission lines, which indicates all cables are properly connected. Standing wave features from the Fourier transform for the antenna are scarcely visible, which corresponds to the length of the stripped part ($\sim$20 mm). Given the huge impedance mismatch of the stripped cable, the standing waves created for the antenna’s transmission line are negligibly small. We ascribe this suppression of standing waves to the effects of a 50-Ohm terminator, which is further supported by the antenna test shown in Fig.~\ref{fig:antenna}(b). Employing the measured transfer functions, we are able to apply RF voltages with constant amplitudes at the tunnel junction throughout a large frequency range (5--40 GHz) by adjusting the power output of the generator. Once calibrated after an iterative optimization procedure of transmission measurements, we observed no significant change in the transfer function for at least several weeks (even possibly several months if there is no change of cables), which implies that the RF cables are well thermalized to fixed temperature points, and that the temperature of these points is not dependent on the LHe level in the cryostat. 

\subsection{Single atom ESR spectroscopy}

\begin{figure}
\includegraphics[width=85mm]{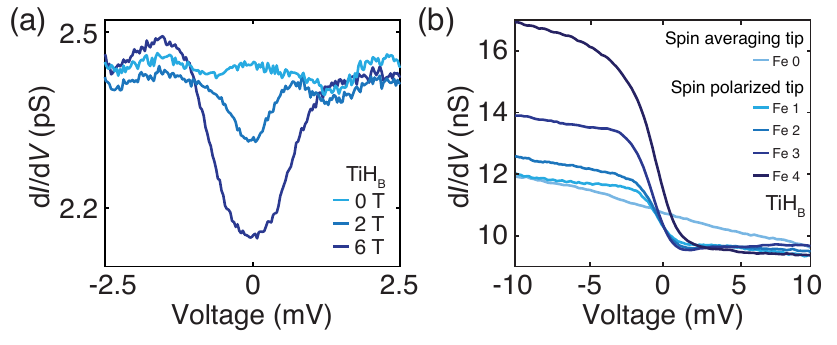}
\caption{\label{fig:expsetup} Characterization of the TiH and the spin-polarized tip for ESR measurement. (a) d$I$/d$V$ spectra of TiH\textsubscript{B} at different magnetic fields ($B_{\mathrm{ext}}=$0, 2, and 6 T in the in-plane direction, $V_{\mathrm{DC}}$~=~4~mV, $I$ = 70 pA, $V_{\mathrm{mod}}$ = 50 $\mu$V, $T$ = 1.3 K). (b) d$I$/d$V$ spectra of TiH\textsubscript{B} with different spin polarization of the tip. The spin-averaging tip corresponds to the Ag-coated STM tip. The STM tip is spin-polarized by picking up Fe atoms on the surface. Each spectrum was measured after picking up Fe atom one-by-one (5 spectra measured with 0 (spin-averaging tip, light blue) to 4 Fe atoms (spin-polarized tip, dark blue) at the tip apex) ($B_{\mathrm{ext}} = 0.9$ T, in-plane direction, $V_{\mathrm{DC}}$ = 10 mV, $I$ = 100 pA, $V_{\mathrm{mod}}$ = 1 mV, $T$ = 1.3 K).}
\end{figure}

We implement the frequency-sweep ESR spectroscopy on a single TiH by applying RF voltages through the tip or the antenna. TiH is a well-known spin-1/2 system\cite{Baeeaau4159,doi:10.1103/PhysRevLett.122.227203}. In a typical spectroscopy of STM, the energy splitting between two spin states of TiH by Zeeman energy is only visible if large external magnetic fields are applied [Fig.~\ref{fig:expsetup}(a)] since the spectral resolution is restricted by the thermal broadening of the tunneling electrons’ energy and electrical noises. The spectral resolution can be significantly enhanced by ESR spectroscopy in STM. In ESR-STM, the energy is measured with respect to the frequency of an applied RF field and thus the energy resolution is limited only by the spin dynamics of the objects such as spin relaxation and coherence times\cite{https://doi.org/10.1002/adma.202107534}. For the ESR measurement, a spin-polarized tip is prepared by picking up several Fe atoms to the tip (typically 3--10 Fe atoms) and checking its polarization by d$I$/d$V$ curves of TiH\textsubscript{O} or TiH\textsubscript{B} that show an asymmetric feature near the zero bias voltage [Fig.~\ref{fig:expsetup}(b)]. Larger spin polarization facilitates ESR measurement but does not guarantee ESR signals since detection and driving of ESR require spin-polarization of the tip and simultaneously inhomogeneous transverse (or longitudinal depending on transitions) magnetic fields from the tip \cite{Baeeaau4159,doi:10.1103/PhysRevLett.122.227203,DOI:10.1126/sciadv.abc5511}.

\begin{figure}
\includegraphics[width=85mm]{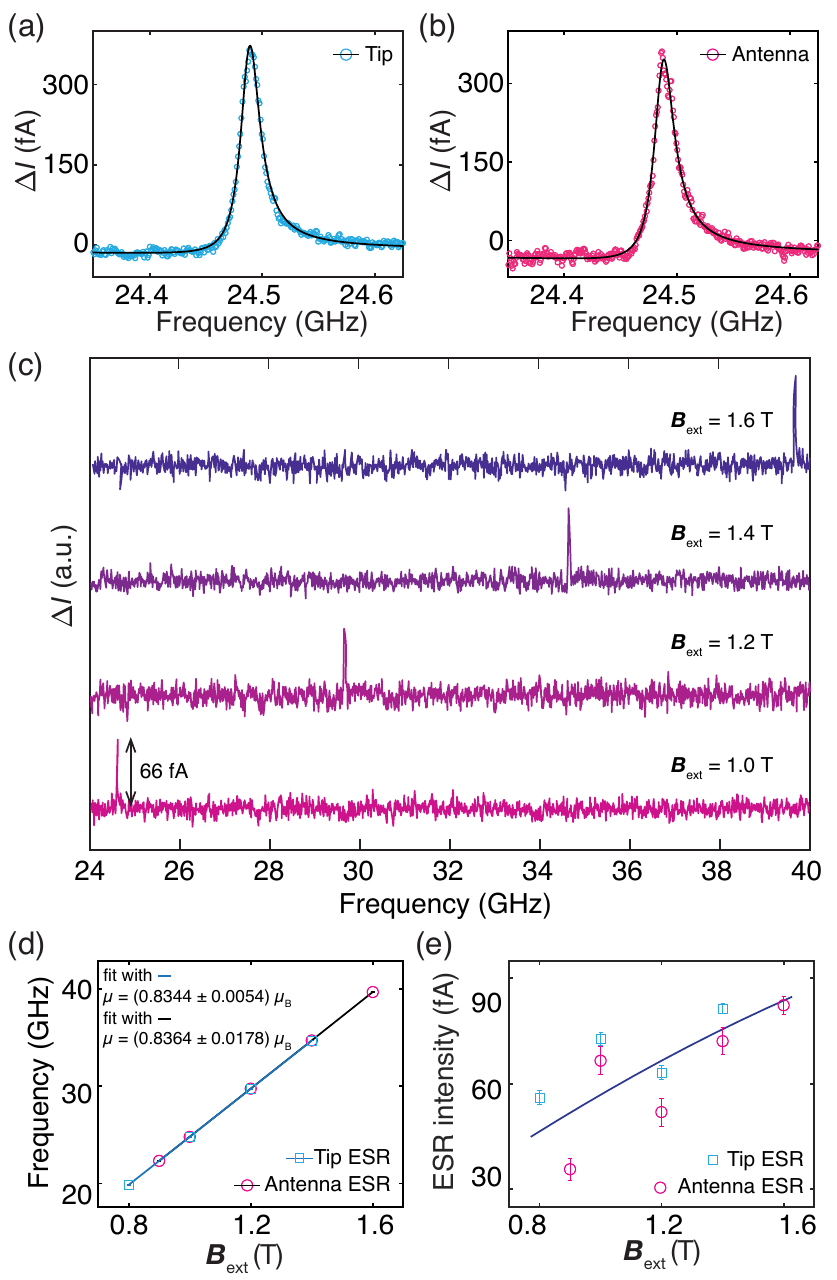}
\caption{\label{fig:ESR1} Comparison of ESR spectra measured using the tip and the antenna for RF transmission. (a) ESR of TiH\textsubscript{B} measured using the tip for RF transmission ($V_{\mathrm{DC}}$~=~50~mV, $I$ = 20 pA, $V_{\mathrm{RF}}$ = 21 mV, $B_{\mathrm{ext}}$ = 1 T, $T$ = 1.3 K). (b) ESR spectrum of the same TiH\textsubscript{B} atom measured using the same microtip but the antenna for RF transmission ($V_{\mathrm{DC}}$ = 50 mV, $I$ = 20 pA, $V_{\mathrm{RF}}$ = 22 mV, $B_{\mathrm{ext}}$ = 1 T, and $T$ = 1.3 K). (c) ESR spectra of TiH\textsubscript{B} at different magnetic fields measured using the antenna. Curves are shifted for clarity ($V_{\mathrm{DC}}$ = 50 mV, $I$ = 20 pA, $V_{\mathrm{RF}}$ = 20 mV, and $T$ = 1.3 K). (d) Resonance frequencies and (e) intensities of ESR signals at different magnitudes of magnetic fields. The values are extracted from the fit to eq. \ref{eq:Fano} of the data in (c) (Antenna; $V_{\mathrm{DC}}$ = 50 mV, $I$ = 20 pA, $V_{\mathrm{RF}}$ = 20 mV, and $T$ = 1.3 K) (Tip; $V_{\mathrm{DC}}$ = 50 mV, $I$ = 20 pA, $V_{\mathrm{RF}}$ = 21 mV, and $T$ = 1.3 K). The solid lines in (d) are linear fits, which gives the TiH\textsubscript{B} magnetic moments, $\mu$. The solid line in (e) corresponds to the estimated ESR intensity based on the Boltzman distribution of spin state populations.}
\end{figure}

We compare the ESR spectra of TiH\textsubscript{B} measured using the tip or the antenna for RF voltage transmission. For RF transmission through the tip, the RF voltages are combined with the DC bias voltage by a bias-tee at RT [Fig.~\ref{fig:wiremap}(b)]. For the ESR measurement using the antenna, we just apply DC bias voltages to the tip, while the RF voltages are applied separately to the antenna. During ESR measurements, transfer functions calibrated for each RF cabling are used to compensate for frequency-dependent RF transmissions. Both measurements use the same signal detection scheme, which detects RF-induced current using a lock-in amplifier while modulating RF power\cite{doi:10.1063/1.4955446}. Figures~\ref{fig:ESR1}(a) and (b) show ESR spectra obtained by applying RF voltages to the tip and the antenna, respectively. Note that we used the same spin-polarized tip and atomic species to measure the change in dc current ($\Delta I$) caused by the population change of Zeeman states and the precession of the transverse magnetization at the resonance frequency\cite{Baeeaau4159}. The ESR spectra are fit to the equation\cite{Baeeaau4159}:
\begin{equation}
    \Delta I =I\textsubscript{0} +  I\textsubscript{p}\cdot \frac{1+\alpha\delta}{1+{\delta}^2}
\label{eq:Fano}
\end{equation}
with $\alpha$ asymmetry factor, $I\textsubscript{p}$ ESR intensity, and $\delta$ normalized frequency ($\delta = \Delta f/{\pi\Gamma}$, where $\Delta f$ is the difference of RF frequency and resonance frequency and $\Gamma$ is the full width at half-maximum of the signal). As analyzed using eq. \ref{eq:Fano}, with well-calibrated transfer functions in two different schemes, we observed ESR spectra with comparable ESR intensity, line-shape, and resonance frequency, which reflects that the underlying mechanism of ESR driving and detection is the same for these two different methods of ESR spectroscopy in STM. 

The ESR signal intensity increases with increasing the population differences between two states. The thermal population can be controlled by changing temperature or energy splitting, where the Zeeman splitting stems from the applied magnetic fields. As shown in Fig.~\ref{fig:ESR1}(c), the ESR spectra were measured at different magnetic fields using the antenna. We observed no spurious features in a large frequency window (24--40 GHz) which might be caused due to miscalibrations of the transfer function. With increasing magnetic field, the resonance frequency of ESR signals linearly increases [Fig.~\ref{fig:ESR1}(c) and (d)], where the slope reflects the magnetic moments of TiH\textsubscript{B}, $\mu = \sim 0.83 \mu_{\mathrm{B}}$, in the in-plane field\cite{10.1103/PhysRevB.104.174408,Baeeaau4159}. With increasing resonance frequencies and thus Zeeman energies, the signal intensity tends to increase (the solid lines in Fig.~\ref{fig:ESR1}(e) show the expected ESR intensities based on the Boltzmann distribution).

\subsection{ESR spectroscopy at variable temperatures}

\begin{figure}
\includegraphics[width=85mm]{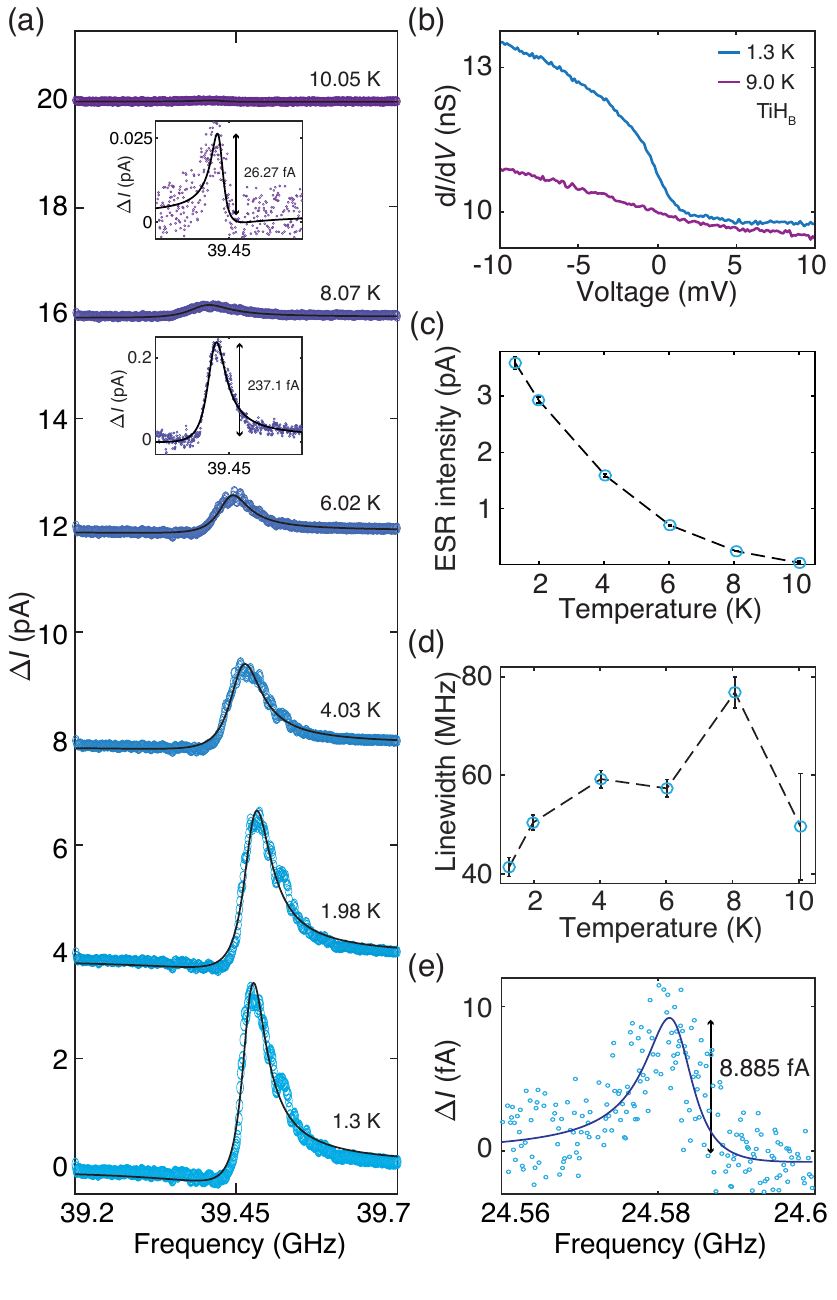}
\caption{\label{fig:ESR2} Temperature dependence of ESR signals. (a) ESR signals of TiH\textsubscript{B} measured at different temperatures from 1.3 K to 10 K ($V_{\mathrm{DC}}$ = 40 mV, $I$ = 40 pA, $V_{\mathrm{RF}}$ = 20 mV, and $B_{\mathrm{ext}}$ = 1.59 T). (b) d$I$/d$V$ spectra measured at different temperatures ($V_{\mathrm{DC}}$ = 10 mV, $I$ = 100 pA, $V_{\mathrm{mod}}$ = 1 mV, and $B_{\mathrm{ext}}$ = 1.59 T). The spectra were measured using the same tip and atom with (a). The assymetric step at zero bias is observable at 1.3 K, while the feature is indistinguishable at 9 K. (c) The ESR signal intensities and (d) linewidths for ESR spectra in (a), as extracted using eq.~\ref{eq:Fano}. (e) The smallest ESR signal measured using our ESR-STM system ($V_{\mathrm{DC}}$ = 32 mV, $I$ = 0.8 pA, $V_{\mathrm{RF}}$ = 30 mV, and $B_{\mathrm{ext}}$ = 1 T, and $T$ = 1.3 K)}
\end{figure}

Taking advantage of large transmission through the antenna at high frequencies, we implement the ESR spectroscopy at higher temperatures which have not been explored before in ESR-STM. We maximize the population difference by increasing the Zeeman energy splitting so that the ESR signals appear at the largest frequency ($\sim$40 GHz) in our frequency window. Figure~\ref{fig:ESR2}(a) shows the ESR spectra at different temperatures. Our STM enables us to conduct experiments at elevated temperatures in the range from 1.3 K to at least 10 K. Using a spin-polarized tip, which shows very strong ESR signals of 3 pA at 1.3 K (out of 40 pA setpoint current), we succeeded in detecting the ESR signal at 10 K, which is $\sim$10 times weaker than the one at 1.3 K. As shown in Fig.~\ref{fig:ESR2}(c), the ESR signal intensity falls exponentially with rising temperature, which is consistent with the exponential dependence of Boltzmann distribution on temperature. While the peak intensity varies dramatically with temperature, there is no discernible trend in the ESR peak broadening [Fig.~\ref{fig:ESR2}(d)], which indicates that there is no major change in the energy resolution of ESR in STM with temperature. In comparison, the spin-dependent feature at zero bias is smeared out in the d$I$/d$V$ curve [Fig.~\ref{fig:ESR2}(b)] since the spectral resolution of the d$I$/d$V$ spectroscopy is limited by the thermal broadening of tunneling electrons’ energy. The strength of ESR spectroscopy in STM, thus, becomes more evident for experiments at higher temperatures. In addition, the highly stable operation of ESR-STM allows for the extremely sensitive measurement of single-spin ESR spectroscopy with the corresponding change in tunnel current by sub-fA [Fig.~\ref{fig:ESR2}(e)]. 

\section{Conclusion}
To summarize, we have presented the design and performance of ESR-STM operating at temperatures of 1--10 K. We for the first time demonstrated ESR of a single atom at elevated temperatures up to 10 K, which was realized by introducing an antenna positioned parallel to the sample and optimizing the operation of ESR-STM. We demonstrated the characterization of the newly introduced antenna in comparison with the early design RF cablings, which provides varieties in designs of RF antennas for integrating ESR in STM and easy modifications of existing STMs for ESR applications. A planned future upgrade will replace the \textsuperscript{4}He gas in the JT refrigerator with \textsuperscript{3}He gas, which should enable stable performance below 0.5 K.

\begin{acknowledgments}
We thank Alexander F. Otte and Alexandre Artaud for their initiative works and fruitful discussions on COMSOL simulations. All authors acknowledge
support from the Institute for Basic Science (IBS-R027-D1).
\end{acknowledgments}

\section*{Data Availability Statement}
The data that support the findings of this study are available from the corresponding author upon reasonable request.

\bibliography{EveRef_220112}

\end{document}